\documentstyle[aps,psfig]{revtex}

\begin{document}

\title{On the Dyadosphere of Black Holes}
\author{Remo Ruffini}
\address{I.C.R.A.-International Center for Relativistic Astrophysics and
Physics Department, University of Rome ``La Sapienza", 00185 Rome, Italy}

\maketitle

\begin{abstract}
The ``dyadosphere" (from the Greek word {\it duas-duados} for pairs)
is here defined as the region outside the horizon  of a
black hole endowed with an electromagnetic field (abbreviated to EMBH
for ``electromagnetic black hole") where the electromagnetic field
exceeds the critical value, predicted by Heisenberg and Euler\cite{[1]} for $e^+
e^-$ pair production. In a very short time ($\sim O({\hbar\over mc^2})$),
a very large number of pairs is created there. I give limits on the EMBH
parameters leading to a Dyadosphere for $10M_{\odot}$ and
$10^5M_{\odot}$ EMBH's, and give as well the pair densities as functions of
the radial coordinate. These data give the initial conditions for the analysis of an enormous pair-electromagnetic-pulse or ``PEM-pulse" which
naturally leads to relativistic expansion. Basic energy requirements for
gamma ray bursts (GRB), including GRB971214 recently observed at $z=3.4$,
can be accounted for by processes occurring in the dyadosphere. 
\end{abstract}
 
\vglue 1cm

\section{Introduction}

The title of the original talk I was supposed to give was"on the critical mass". I was planning to cover the latest developments since my early work on the role of the critical mass in the physics and astrophysics of neutron stars and Black Holes\cite{[2]}. To discuss as well, how after my propoposal of identifying Cygnus X1 as a Black Hole\cite{[3]} approximately ten binary systems with very similar characteristic have been found in our own galaxy. 

I was also going to stress that very little progress has been made in the intervening years about the determination of the critical mass of neutron stars due to the formidable theoretical difficulties in understanding the behaviour of matter at supranuclear densities and that indeed the theorem about the absolute upper limit to a neutron star mass, established by C. Rhoades and myself\cite{[4]}, stands still as the pivotal point in the paradigm of identification of Black Holes versus Neutron Stars. 

It was my intention to show how, surprisingly, progress is being made recently in gaining a new  understanding of the white dwarf critical mass: a classic problem considered by many people well settled, which is extremely attractive from a theoretical point of view. The white dwarfs offer the possibility of testing the concept of critical mass in a regime where all physical theories work in the exact range of their tested validities: there are not unknown due to extrapolation of physical theories to yet unexplored domains as in the matter at supranuclear densities. The new effects can be computed by generalizing to a relativistic treatment the classical results of the Feynmann-Metropolis- Teller work based on the Fermi-Thomas model of the atom to relativistic regimes. The consequences on the estimate of the critical mass can be of paramount importance in the understanding the observed values of the neutron stars in binary systems as well the process of formation of such systems.

However, the recent observations of the Beppo-SAX satellite, the discovery of a
very regular afterglow to GRB's, the fact their x-ray flux varies
regularly with time according to precise power laws (Costa {\it et al.}\cite{[5]}; 
Van Paradijs {\it et al.}\cite{[6]}), and especially the optical and infrared identification of these sources
(Kulkarni {\it et al.}\cite{[7]}, Halpern {\it et al.}\cite{[8]} and Ramaprakash {\it et al.} \cite{[9]}) which has established their correct cosmological setting and
determined their formidable energy requirements
convinced me to reconsider some theoretical results on black hole
astrophysics (see e.g. Ruffini {\it et al.}\cite{[10]}, Fang, Ruffini and Sato\cite{[11]}) and to develop a detailed
model for a direct confrontation with the observational results. 

For these reasons I have deceided, with the consensus of the organizers, to modify slightly the title of my talk to a new talk and deceided to insert this material on the critical mass in a different publication\cite{[12]}. This will allow me to introduce today a most urgent new concept: the Dyadosphere of a Black Hole and offer this new concept to Humitaka Sato for his sixtyfifth birthday.

\section{The basic results on EMBH}

It is now generally accepted that Black Holes are uniquely characterized by their total 
mass-energy $E$, charge $Q$, and angular momentum $L$\cite{[13]}. The uniqueness theorem has finally been proved after some twenty five years of mathematical work by numerous authors recently reviewed in ref.\cite{[14]}. 

I {\it assume} that the process of gravitational collapse for a 
core larger than the neutron star critical mass will generally lead to
a black hole characterized by {\it all} three fundamental parameters of
mass-energy $M$, angular momentum $L$, and charge $Q$ (see Ruffini and Wheeler \cite{[13]}). While the first two parameters $M$ and $L$  evolve on very long
time scales, the charge $Q$ will dissipate on much shorter time scales. In
particular I show that if the electromagnetic field of an EMBH is
overcritical in the sense of Heisenberg and Euler\cite{[1]}, it will
dissipate on an extremely short time scale of approximately $10^7$ 
sec.~through a new
process involving the ``dyadosphere" of a black hole. In a very short
time $\sim O({\hbar\over mc^2})$, a very large number of pairs is created
there and reaches thermodynamic equilibrium with a photon gas. In
the ensuing enormous  PEM-pulse, a large fraction of the extractable
energy of the EMBH in the sense of Christodoulou-Ruffini\cite{[15]} will be 
carried away. The PEM-pulse will interact with some of the
baryonic matter of the uncollapsed material and  the associated emission
is closely related to the observed properties of GRB sources. The
electromagnetic field of the remnant will further dissipate in the
acceleration of cosmic rays or in the propulsion of jets on much longer
time scales. 

The  Christodoulou-Ruffini\cite{[15]} energy-mass formula
for black holes gives
\begin{eqnarray}
E^2&=&M^2c^4=\left(M_{\rm ir}c^2 + {Q^2\over2\rho_+}\right)^2+{L^2c^2\over
\rho_+^2},\label{em}\\
S&=& 4\pi\rho_+^2=4\pi(r^2_++{L^2\over c^2M^2})=16\pi \left({G^2\over c^4}\right)M^2_{\rm ir},
\label{s}
\end{eqnarray}
with
\begin{equation}
{1\over\rho_+^4}\left({G^2\over c^8}\right)\left( Q^4+{L^2c^2\over4}\right)\leq 1,
\label{s1}
\end{equation}
where $M_{\rm ir}$ is the irreducible mass, $r_{+}$ is the
horizon radius and $\rho_+$ is the quasi spheroidal cylindical coordinate of the horizon evaluated at the equatorial plane. (I use c.g.s.~units). From eqs.~(\ref{em}) and
(\ref{s1}) it follows that
up to 29$\%$ of the mass-energy of an extreme rotating black hole with
$L_{\rm max}=2r^2_{+}c^3/G$ can be stored as rotational energy and
gedanken experiments have been conceived to extract such energy (see e. g. Ruffini and Wheeler\cite{[13]} and ref.\cite{[16]}). 
In the case of black holes endowed with an electromagnetic field, it follows 
from the same equations that up to 50$\%$ of the mass
energy of an extreme EMBH with $Q_{\rm max}=r_+c^2/\sqrt{G}$ can be stored in its
electromagnetic field. It is appropriate to recall that even in the case
of an extreme EMBH the charge to mass ratio is $\sim 10^{18}$ smaller then
the typical charge to mass ratio found in nuclear matter, owing to the
different strength and range of the nuclear and gravitational
interactions. In other words it is enough to
have a difference of one quantum of charge per $10^{18}$ nucleons in
the collapsing matter for an EMBH to be extreme. 
By applying the classic work of
Heisenberg and Euler\cite{[1]} to EMBH's, as reformulated in a
relativistic-invariant form by Schwinger\cite{[17]},
Damour and Ruffini\cite{[18]} showed that a
large fraction of the energy of an EMBH can be extracted by pair
creation.  This energy extraction process only works for
EMBH black holes with $M_{\rm ir}< 10^6M_{\odot}$. They also claimed
that such an energy source might lead to a natural explanation for GRB's and for ultra high energy cosmic rays.

The general considerations presented in Damour and Ruffini\cite{[18]} are correct. However,
that work has an underlying assumption which only surfaces in the very last formula: that the pair created in the process of vacuum polarization is absorbed by the EMBH. That view is now fundamentally modified here by the introduction of the novel concept of the Dyadosphere of an EMBH
and by the considerations which follow from its introduction.

\section{The extension of the Dyadosphere}

For simplicity we use the nonrotating Reissner-Nordstrom EMBH to illustrate the basic gravitational-electrodynamical process. The rotating Kerr-Newmann EMBH case will be considered elsewhere\cite{[19]}.

By introducing the dimensionless mass and charge parameters $\mu={M\over M_{\odot}}>3.2$, $\xi={Q\over Q_{\rm max}}\le 1$, the horizon radius may be expressed as
\begin{equation}
r_{+}={GM\over c^2}\left[1+\sqrt{1-{Q^2\over GM^2}}\right]
=1.47\cdot 10^5\mu (1+\sqrt{1-\xi^2})\hskip0.1cm {\rm cm}.
\label{r+}
\end{equation}
Outside the horizon the electromagnetic field measured in the orthonormal tetrad of an observer at rest at a given point $r$ in the usual Boyer-Lindquist coordinates (see e.g. \cite{[20]}) has only one nonvanishing component ${\bf E}= {Q\over r^2}\hat{r}$ 
along the radial direction. We can evaluate the radius $r_{\rm ds}$ at which the electric field strength
reaches the critical value ${\cal E}_{\rm c}={m^2c^3\over\hbar e}$ introduced by Heisenberg and Euler, where
$m$ and $e$ are the mass and charge of the electron.  
This defines the outer radius of the Dyadosphere, which extends down to the horizon and within which the electric field strength exceeds the critical value. Using the Planck charge $q_{\rm c}= (\hbar c)^{1\over2}$ and the Planck mass $m_{\rm p}=({\hbar c\over G})^{1\over2}$, I express this outer radius
in the form
\begin{equation}
r_{\rm ds}=\left({\hbar\over mc}\right)^{1\over2}
\left({GM\over c^2}\right)^{1\over2} 
\left({m_{\rm p}\over m}\right)^{1\over2}
\left({e\over q_{\rm p}}\right)^{1\over2}
\left({Q\over\sqrt{G} M}\right)^{1\over2}
=1.12\cdot 10^8\sqrt{\mu\xi} \hskip0.1cm {\rm cm},
\label{rc}
\end{equation} 
which clearly shows the hybrid gravitational and quantum nature of this quantity. The radial interval $r_{+}\leq
r \leq r_{\rm ds}$ describing the Dyadosphere as a function of the  mass is illustrated in Fig.[1] for selected values of the charge paramter $\xi$. 
It is important to note that the Dyadosphere radius is maximized
for the extreme case $\xi=1$ and that the region exists for EMBH's with mass larger than the upper limit for neutron stars, namely $\sim 3.2M_{\odot}$ all the way up to a maximum mass of $6\cdot 10^5M_{\odot}$.
Correspondingly smaller values of the maximum mass are obtained for values of $\xi=0.1,0.01$ as indicated in this figure. For EMBH's with mass larger than the maximum value stated above, the electromagnetic field (whose strength decreases inversely with the mass) never becomes critical.

\section{The inner structure of Dyadosphere}

I turn now to the crucial issue of the number and energy densities of pairs created in the Dyadosphere. If we consider a shell of proper thickness $\delta\ll {MG\over c^2}$, the electric field is aproximately constant. We can then at each value of the radius $r$ model the electric field as created by a capacitor of width $\delta$ and surface charge density 
\begin{equation}
\sigma(r)={Q\over 4\pi r^2},
\label{sq}
\end{equation}
If we now turn to the process of pair creation in order to apply the QED results, we assume $\delta={\hbar\over mc}$ and we can express the rate of pair creation at a given radius $r$ by \cite{[1],[18]}
\begin{equation}
{dN\over\sqrt{-g} d^4x}= {1\over4\pi c}\left({eE\over\pi\hbar}\right)^2
e^{-{\pi E_c\over E}}={1\over4\pi c}\left({4e\sigma\over\hbar}\right)^2
e^{-{\pi \sigma_c\over \sigma}},
\label{rate1}
\end{equation}
where $E=4\pi\sigma$, $\sigma_c={1\over4\pi }{\cal E}_{\rm c}$ is the critical surface charge density and $\sqrt{-g} d^4x$ is 
the invariant four volume. We have for each value of the radius $r$, the rate of pair creation per unit proper time
\begin{equation}
{dN\over d\tau}= {1\over4\pi c}\left({4e\sigma\over\hbar}\right)^2
e^{-{\pi \sigma_c\over \sigma}}4\pi r^2\left({\hbar\over mc}\right).
\label{rate2}
\end{equation}

The pair creation process will continue until a value of the field $E\simeq E_c$ is reached, or correspondingly, in the capacitor language, till the surface charge density reaches the critical value $\sigma_c$. For $E<E_c$ or $\sigma<\sigma_c$ the pair creation process is exponentially supressed. We then have
\begin{equation}
\sigma-\sigma_c ={e\over4\pi r^2}{\Delta N\over \Delta \tau}\Delta \tau.
\label{c}
\end{equation}
Based on eq.(\ref{rate2}), one can get,
\begin{equation}
\Delta \tau= {\sigma-\sigma_c\over {e\over4\pi c}\left({4e\sigma\over\hbar}\right)^2
e^{-{\pi \sigma_c\over \sigma}}\left({\hbar\over mc}\right)}
< 1.99\left({\hbar\over mc^2\alpha}\right)
=1.7610^{-19}{\rm sec.},
\label{dis}
\end{equation}
where $\alpha={e^2\over4\pi\hbar c}$ is the fine structure constant. ( Details are given in ref.\cite{[21]}.) The time given by eq.(\ref{dis}) is so short, that the light travel time is smaller or aproximately equal to the width $\delta$. Under these circumstances the correlation between shells can be approximately neglected, thus we can
justify the aproximation of describing the pair creation process shell by shell.

If we now turn to the Dyadosphere its extension goes from the horizon $r_+$ all the way to a radius $r_{ds}$ where ${Q\over4\pi r_{ds}}=E_c$. The inner layer of the first shell of width $\delta$, consisting of charges opposite to the one of EMBH's, will be captured by the horizon r+, leading to a new EMBH with charge $Q_c= 4\pi r_+^2E_c$. The 
outer layer, of oppositely charged particles, will enter the Dyadosphere. The remaining $\sim (r_{ds}-r_+)/{\hbar\over mc}$ shells will also contribute to the plasma constituting the Dyadosphere. It is noteworty to stress tha the number of pair created is {\it not}, as nively expected $N_{e^+e^-}={Q-Q_c\over e}$, but the number is tremendously amplified. In the limit 
$r_{ds}\gg {GM\over c^2}$, we have

\begin{equation}
N_{e^+e^-}\simeq {Q-Q_c\over e}\left[1+{
(r_{ds}-r_+)\over {\hbar\over mc}}\right].
\label{n}
\end{equation}

\section{Density and energy density of pairs in the Dyadosphere}

We can now estimate from eq.(\ref{c}) the number of pairs created at a given radius $r$ in a shell volume of proper thickness $\delta={\hbar\over mc}$
\begin{equation}
\Delta N={4\pi r^2\over e}(\sigma-\sigma_c)={Q\over e}\left[1-\left({r\over r_{ds}}\right)^2\right],
\label{nn}
\end{equation}
and correspondingly the number density of pairs created as a function of the radial coordinate, details in ref.\cite{[22]} and \cite{[23]}
\begin{equation}
n_{e^+e^-}(r)={\Delta N\over4\pi r^2\left({\hbar\over mc}\right)}
={Q\over e4\pi r^2\left({\hbar\over mc}\right)}\left[1-\left({r\over r_{ds}}\right)^2\right].
\label{nd}
\end{equation}

In Figs.[2] and [3], we plot the density of pairs for
$10M_{\odot}$ and $10^5M_{\odot}$ EMBH for selected values of $\xi$. These two values of the mass were chosen to be representative of objects typical of the galactic population or for the nuclei of galaxies compatible with our upper limit of the maximum mass of $6\cdot 10^5M_{\odot}$. 

Knowing the initial electrostatic energy density, and the final electric field value  
$E_{\rm c}(r)={Q_{\rm c}\over r^2}$ holding after the evolution of the Dyadosphere, we compute the energy density of created pairs as a function of the radial coordinate by evaluating 
difference between the initial and final field configurations in the Dyadosphere
\begin{equation}
\rho_{e^+e^-}={1\over8\pi}(E^2(r)-E_c^2(r))\ .
\label{es}
\end{equation}  
Their total energy is then
\begin{equation}
E^{\rm tot}_{e^+e^-}={1\over2}{Q^2\over r_+}(1-{r_+\over r_{\rm ds}})(1-
\left({r_+\over r_{\rm ds}}\right)^2).
\label{te}
\end{equation}

By taking the ratio of eq.(\ref{es}) and eq.(\ref{nd}), we obtain the average energy per pair at each value of the radial coordinate.  Fig.[4] plots this quantity for an
EMBH of $10M_{\odot}$ and Fig.[5] for an EMBH of $10^5M_{\odot}$.
In the first case the energy of pairs near the horizon can reach $10$GeV, while in the second case it never goes over a few MeV. 
These two values of the mass were chosen to be representative of objects typical of the galactic population or for the nuclei of galaxies compatible with our upper limit of the maximum mass of $6\cdot 10^5M_{\odot}$. 

Due to the very large pair density given by eq.[13]
and to the sizes of the
cross-sections for the process $e^+e^-\leftrightarrow \gamma+\gamma$, 
the system is expected to thermalize to a plasma configuration for which
\begin{equation}
N_{e^+}=N_{e^-}=N_{\gamma}=N_{\rm pair}
\label{plasma}
\end{equation}
and reach an average temperature
\begin{equation}
kT_\circ={ E^{\rm tot}_{e^+e^-}\over3N_{\rm pair}\cdot2.7},
\label{t}
\end{equation}
where $k$ is Boltzmann's constant.
The average energy per pair ${ E^{\rm tot}_{e^+e^-}\over N_{\rm pair}}$ is shown as a function
of the EMBH mass for selected values of the charge parameter $\xi$  in
Fig.[6].

\section{The energy extraction from EMBH and the Gamma-Ray bursts afterglow}

Finally we can now estimate the total energy extracted by the pair creation process in EMBH's of
different masses for selected values of the charge parameter $\xi$ and compare and contrast these values with the
maximum extractable energy given by the mass formula for Black Holes (see eqs.~(\ref{em}) and 
(\ref{s1}). This comparison is summarized in Fig.[7]. The efficiency of energy extraction by pair creation sharply decreases as the maximum value of the EMBH mass for which vacuum polarization occurs is reached. In the opposite limit the efficiency approaches $100\%$.

As shown by Ruffini {\it et al.}\cite{[24]} the further evolution of
this plasma leads to a relativistic expansion, $e^+ e^-$ annihilation and
an enormous pair-electromagnetic-pulse ``PEM-pulse". 

The evolution of this PEM-pulse has been studied by introducing a
variety of models based on relativistic hydrodynamical
equations, on quasi analitical and numerical treatments and comparing and constracting the results with the ones obtained by numerical simulation in the super-computer of Livermore in California \cite{[24]}. It was possible to follow the evolution of the PEM-pulse and its reaching relativistic velocities. A variety of models has been considered both in the absence and presence of baryonic matter, three examples are given in Fig.8. 

The PEM-pulse in its evolution leads to a variety of emmision profiles, as a function of the EMBH mass, charge and of the amount of baryonic matter surrounding EMBH's. A brief example is given in Fig.9 where most explicitly it is shown how the observed power law the afterglow of the Gamma-rays burst is recovered with their characteristic time scale of $10^7$ sec. Details for the computations and other example are given in ref.\cite{[25]}.

\section{Conclusions}

If this basic scenario is confirmed by observations, other
fundamental questions should be investigated to understand the
origin of the dyadosphere. Some preliminary work along these lines has
already been done by Wilson \cite{[26],[27]} who has shown how relativistic
magneto-hydrodynamical processes
occurring in the accreting material around {\it an already formed black hole}
lead naturally to very effective charge separation and to reaching
the  critical value of the charge given by the limiting value of Eq.~(\ref{s1}) 
on a timescale of $10^2$--$10^3 GM/c^3$.
These studies show the very clear tendency of powerful processes of charge separation and of magnetosphere formation to occur in accretion processes.  They cannot, however, be simply extrapolated to the formation of the dyadosphere.  The time scale of the dyadosphere discharge is of the order of $10^{-19}$ sec (for a detailed discussion including relativistic effects, see Jantzen and Ruffini, \cite{[21]}).  Such a time scale is much shorter than the characteristic magnetohydrodynamical time scales.  We expect that the formation of the dyadosphere should only occur during the gravitational collapse itself and {\it in the process of formation} of the EMBH, with the formation of a charge depleted region with an electric field sufficient to polarize the vacuum.
A complementary aspect dealing  with  the extremization and equipartition of the electromagnetic energy in a gravitating rotating object advanced in Ruffini and Treves \cite{[28]} should also now be considered again. 
Modeling such a problem seems to be extremely difficult at the present time, also in absence of detailed observations narrowing down the values of the fundamental parameters involved.  Confirmation of the basic predictions of the dyadosphere model by observations of gamma ray bursts and their afterglow would provide motivation and essential information to attack such a difficult problem.

The model of GRB's, based on the Dyadosphere, is very different from the ones debated in the last ten years in the scientific literature, it can overcome some of their basic difficulties and on some specific aspects it can also have important analogies:

\begin{itemize}

\item	
the EMBH drastically differs from the most popular binary neutron stars model \cite{[29]} debated in the last 16 years. It presents over those models the distinct advantage of not having the recognized difficulty of explaining the energetic of GRB's \cite{[30]}, and especially offers the possibility of computing explicitly and uniquely the details of the initial conditions by the energetic of the Dyadosphere as given in the above Fig.4 and Fig.5.

\item
the further evolution of the Dyadosphere (the PEM-pulse), has been analyzed  by a variety of theoretical models based on relativistic hydrodynamics equations both on semi- analytical-numerical idealized models as well as with the help of a fully general relativistic hydrodynamical code. The combined results are at variance with many of the considerations in the current published papers dealing with fireballs and relativistic winds \cite{[31]}, though for a few aspects they present striking analogies. Details are given in ref.\cite{[32]}.

\item 
the new treatment lead uniquely to the computation of the further evolution of the PEM Pulse in the afterglow era and again these specific analysis which can be uniquely derived from the above assumptions are at variance with the ones in the existing literature \cite{[33]} though they present some interesting analogies and they appear to be consistent with the latest observations \cite{[25],[32]}. 

\end{itemize}

It goes without saying that the Heisenberg-Euler
process of vacuum polarization considered here and in Damour and Ruffini \cite{[18]}
has nothing to do
with the evaporation of black holes considered by Hawking \cite{[34]}
either from a qualitative or quantitative point of view. The effective
temperature of black hole evaporation given by Hawking for a
$10M_{\odot}$ black hole is $T\sim 6.2\cdot 10^{-9}K^\circ$, which 
implies a black hole lifetime of $\tau\sim 10^{66}$ years 
and an energy flux of $10^{-24}$ergs/sec, while
a $10^5M_{\odot}$ black hole has a Hawking temperature of $T\sim 6.2\cdot
10^{-13}K^\circ$ and lifetime of $\tau\sim 10^{78}$ years with an energy 
flux of $10^{-32}$ergs/sec.

From a fundamental point of view, the process occuring in the Dyadosphere
represents the first mechanism capable of extracting large amounts of energy from a Black Hole with an extremely high efficiency.

\newpage 

\begin{figure}[t]
\centerline{
\psfig{figure=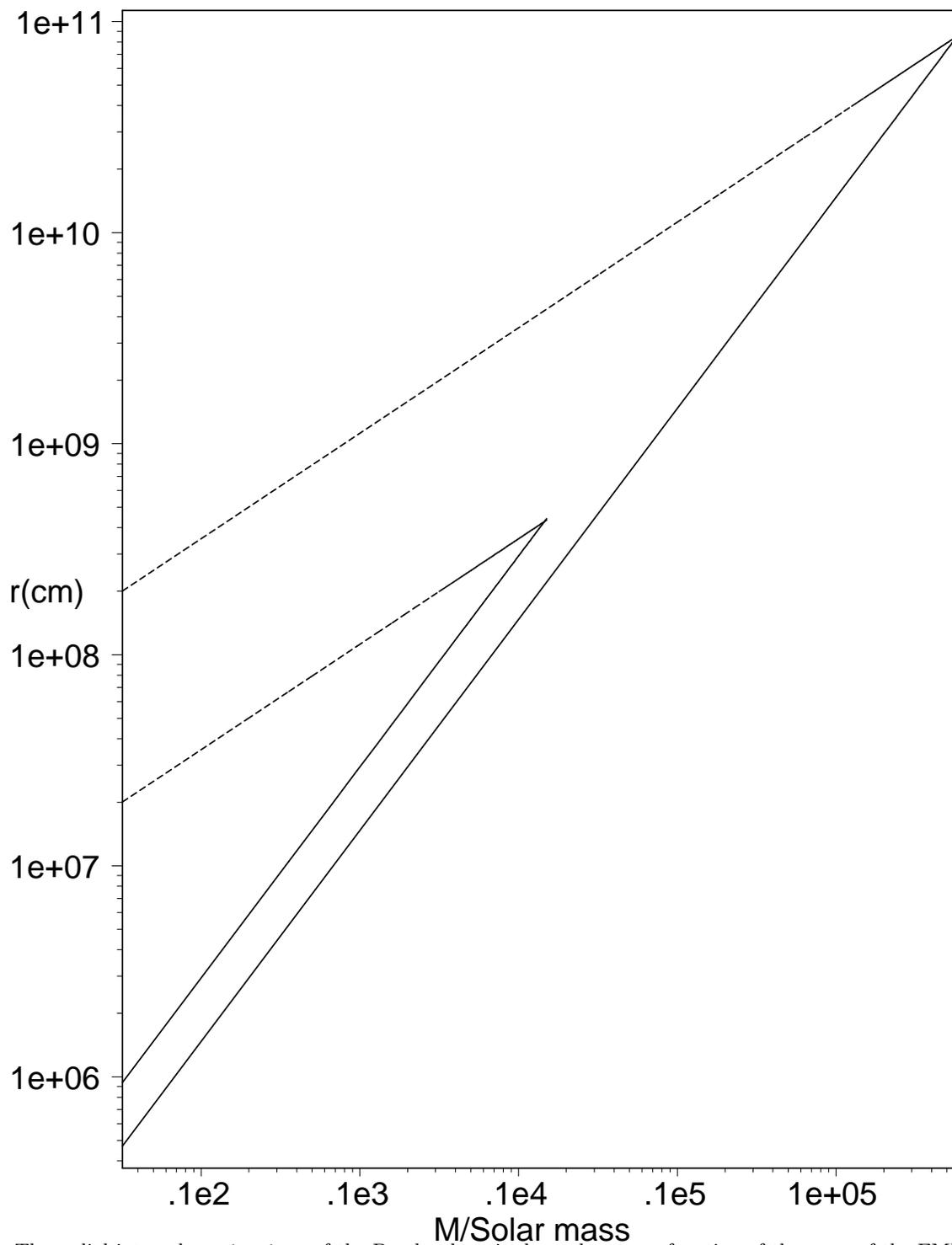}
}
\caption[]{The radial interval $r_+\leq r \leq r_{\rm ds}$ of the Dyadosphere  is shown here as a function of the mass of  the EMBH in solar mass units for the charge parameter values $\xi=1$ (upper curve pair) and $\xi=0.01$ (lower curve pair).
The continuous lines  correspond to the horizon radius $r_+$ given in eq.~(\ref{r+}), and the dotted lines to the Dyadosphere radius $r_{\rm ds}$ given in eq.~(\ref{rc}).}
\label{fig: fig1}
\end{figure}
\newpage 
\begin{figure}[t]
\centerline{
\psfig{figure=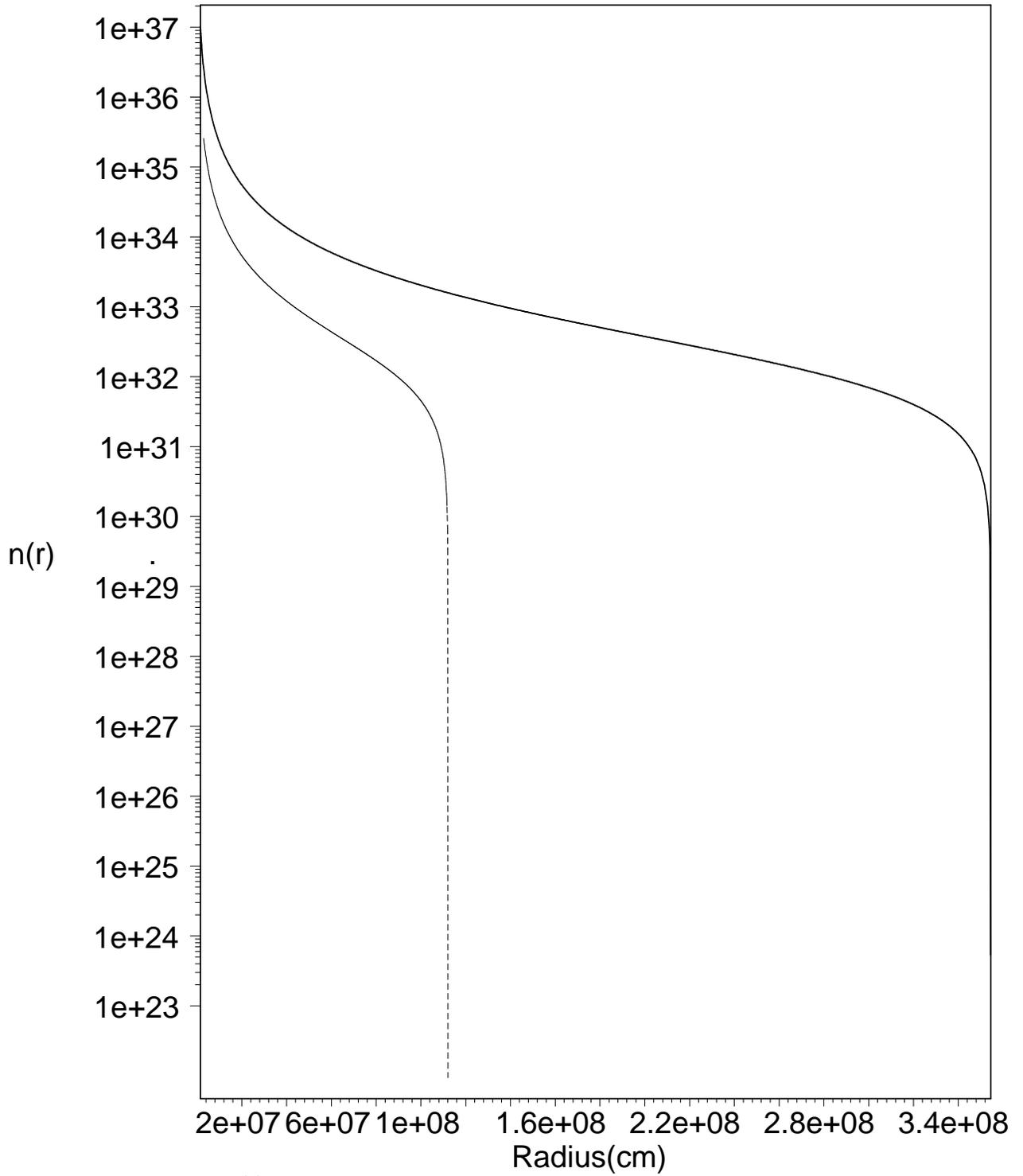}
}
\caption[]{ The density $n_{e^+e^-}(r)$
as a function of the radial coordinate for EMBH's of
$10M_\odot$ corresponding to the charge parameter values 
$\xi=1$ (upper curve) and $\xi=0.1$ (lower curve).
}\label{fig: fig2}
\end{figure}
\newpage
\begin{figure}[t]
\centerline{
\psfig{figure=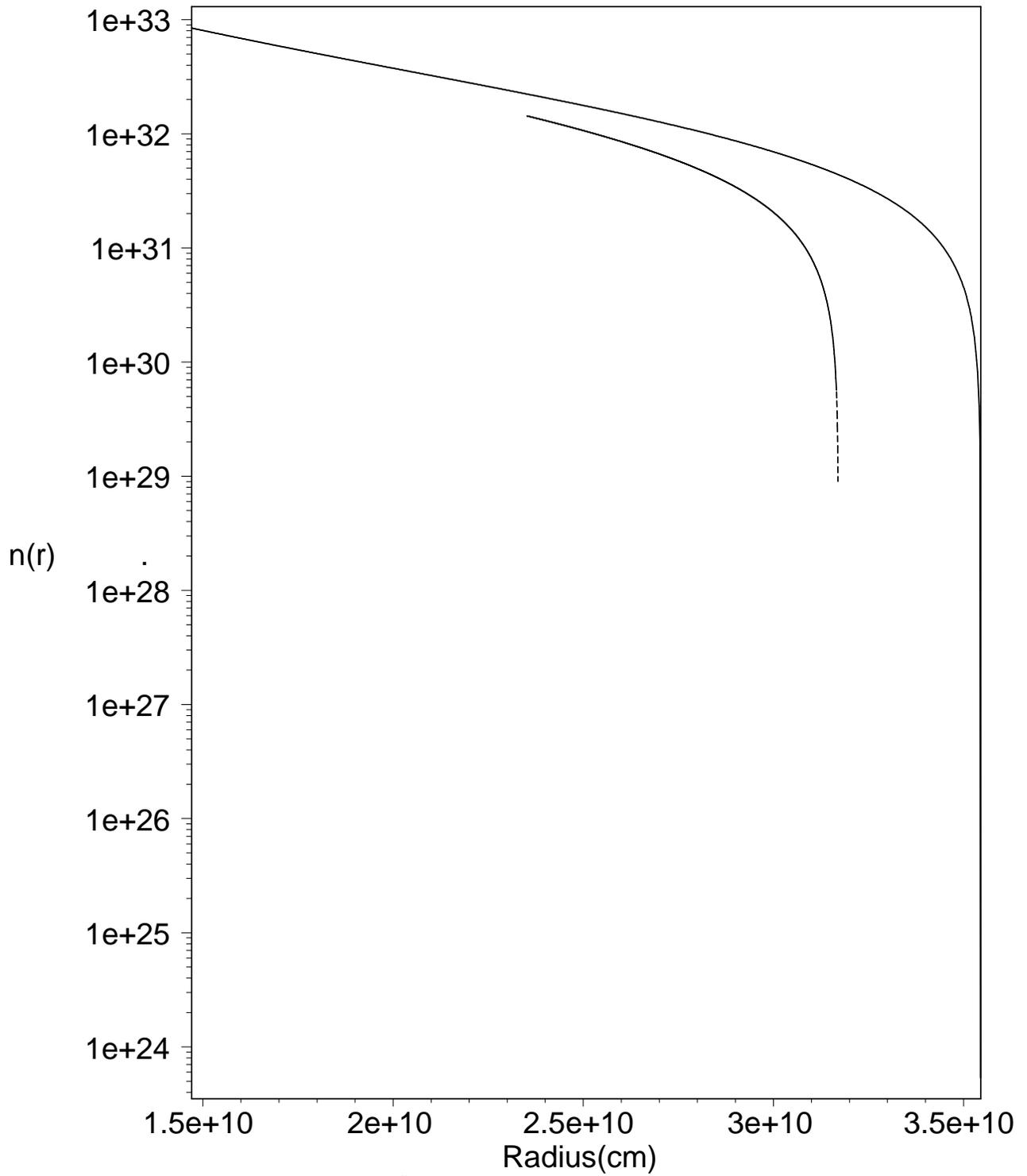}
}
\caption[]{The same as Fig.[2] for an EMBH of
$10^5M_\odot$ corresponding to the charge parameter values 
$\xi=1$ (upper curve) and $\xi=0.8$ (lower curve).}
\label{fig: fig3}
\end{figure}
\newpage
\begin{figure}[t]
\centerline{
\psfig{figure=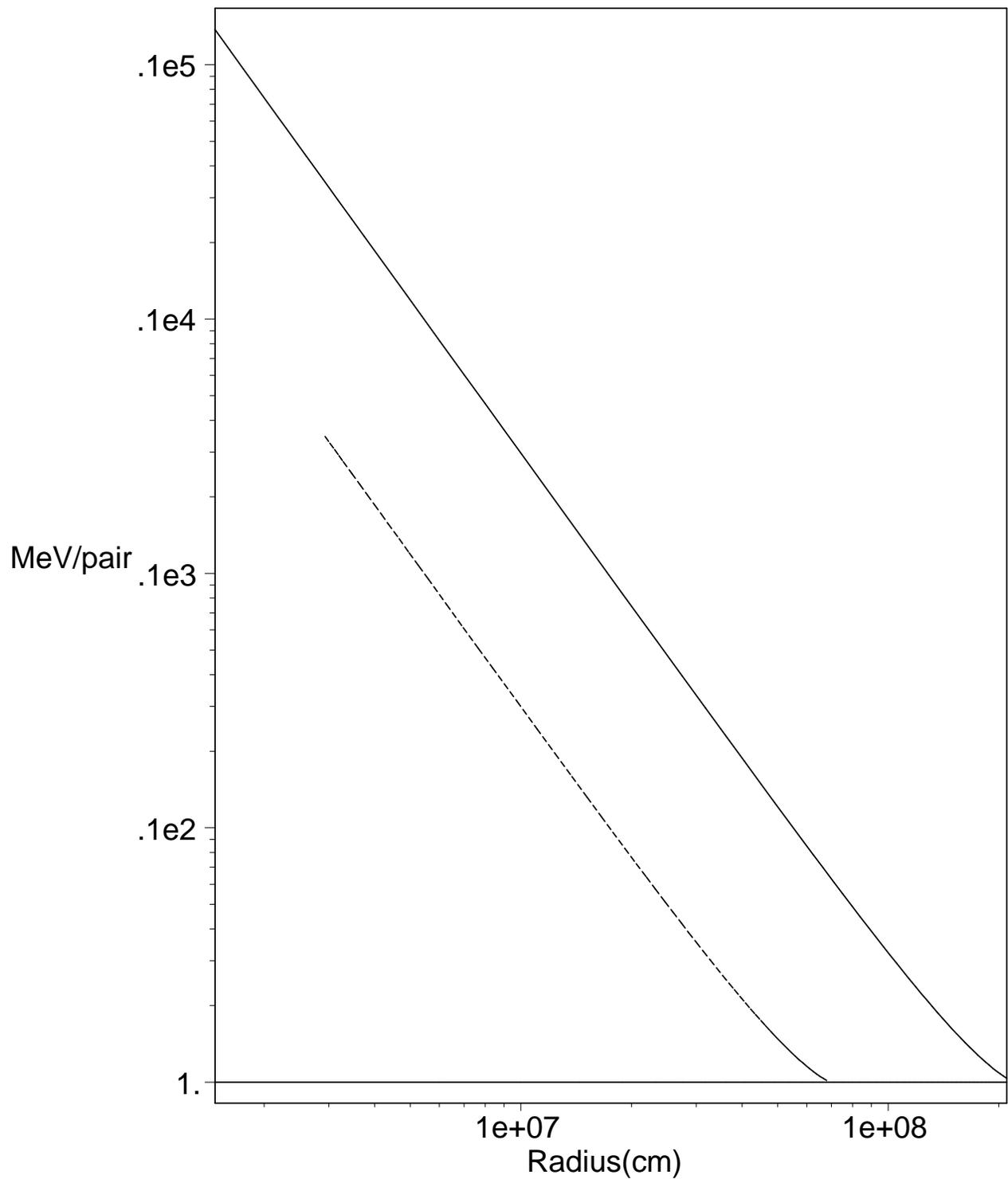}
}
\caption[]{Average energy per pair in MeV plotted as a function of the radius for an EMBH of $10M_\odot$ for the charge parameter values $\xi=1$ (upper curve) and $\xi=0.8$ (lower curve).}
\label{fig: fig4}
\end{figure}
\begin{figure}[t]
\centerline{
\psfig{figure=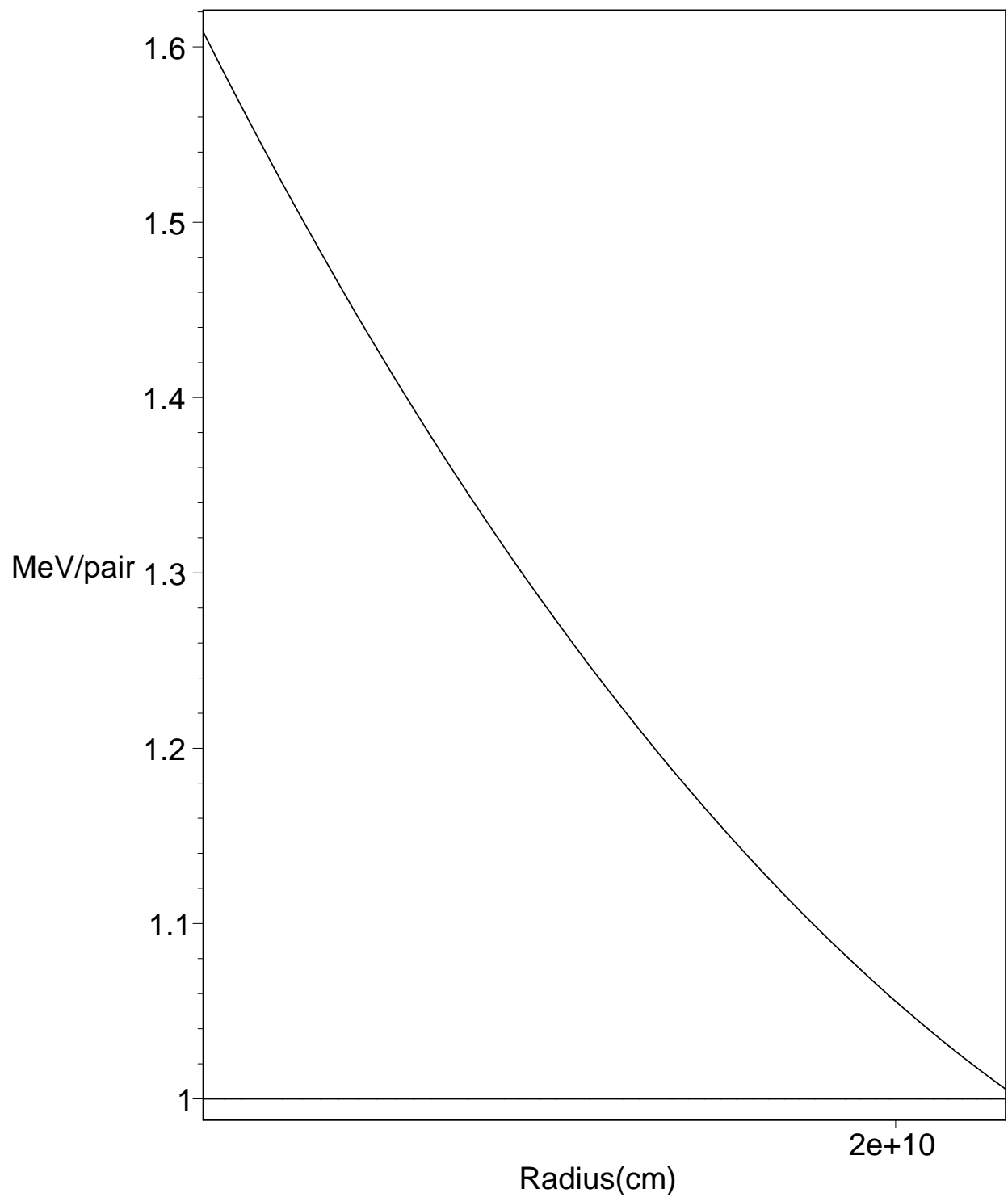}
}
\caption[]{ The same as Fig.~[4] for an EMBH of $10^5M_\odot$ and $\xi=1$.}
\label{fig: fig5}
\end{figure}
\newpage
\begin{figure}[t]
\centerline{
\psfig{figure=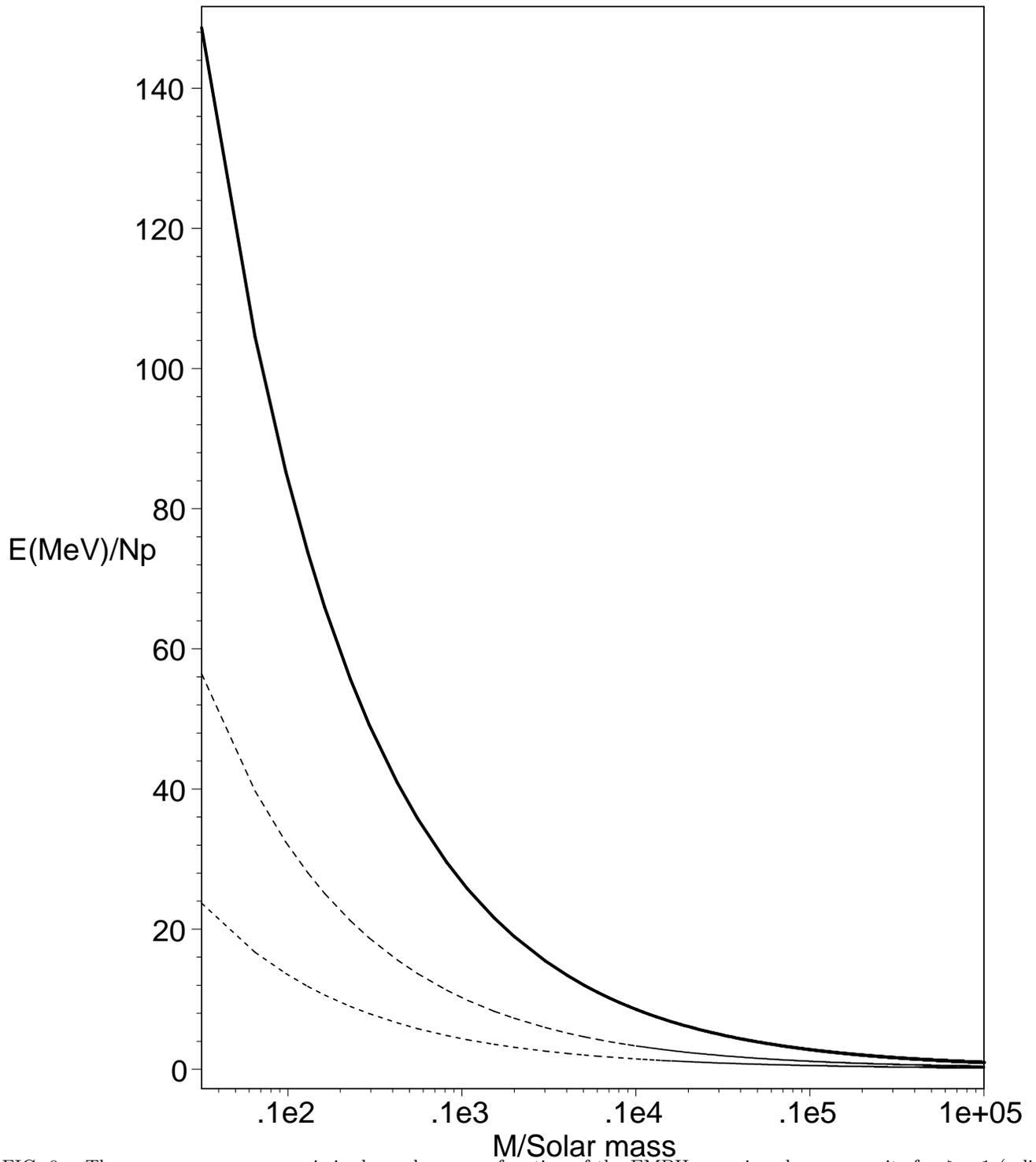}
}
\caption[]{ The average energy per pair is shown here as a function of
the EMBH mass in solar mass units 
for $\xi=1$ (solid line), $\xi=0.5$ (dashed line) and
$\xi=0.1$ (dashed and dotted line).}
\label{fig: fig6}
\end{figure}
\newpage
\begin{figure}[t]
\centerline{
\psfig{figure=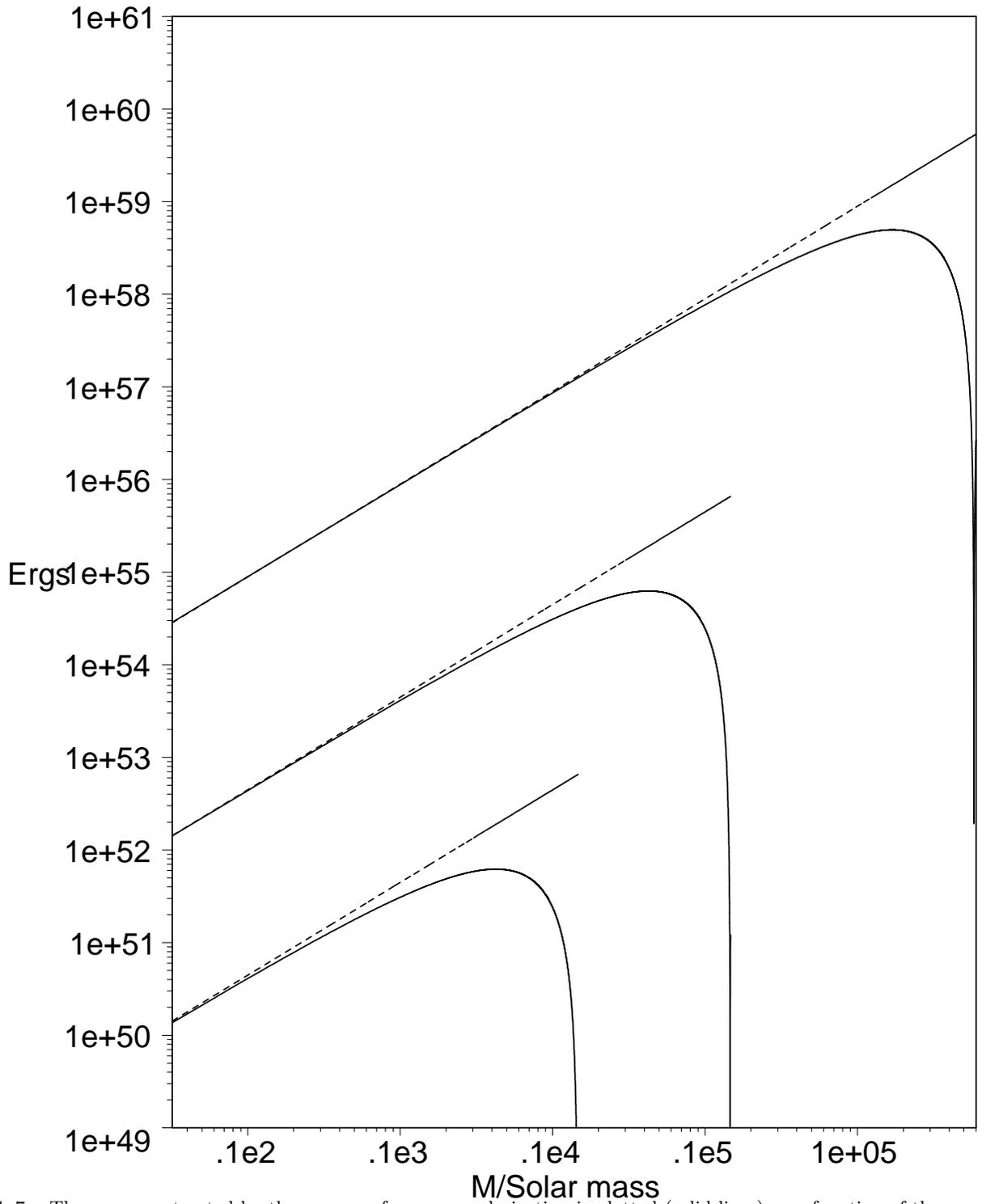}
}
\caption[]{ The energy extracted by the process of vacuum polarization is plotted (solid
lines) as a function of the mass $M$ in solar mass units for selected values of the 
charge parameter $\xi=1, 0.1, 0.01$ for an EMBH. For comparison we have also plotted the maximum energy extractable from an EMBH (dotted lines) given by eq.~(\ref{em}).}
\label{fig: fig7}
\end{figure}
\newpage
\begin{figure}[t]
\centerline{
\psfig{figure=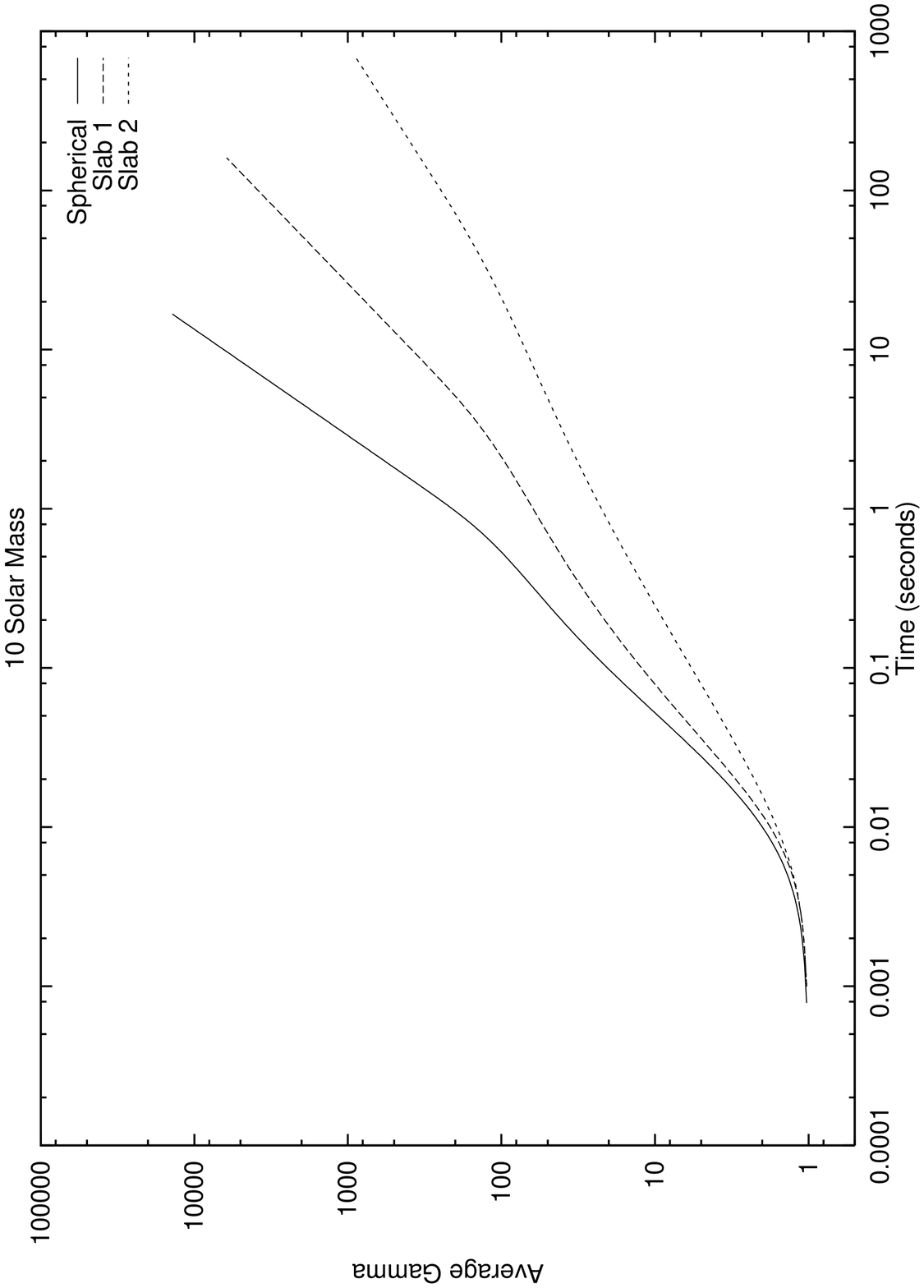}
}
\caption[]{ The relativistic expansion factor ($\Gamma$) for three different analitical models of PEM-pules, in the absence of baryonic matter, are given. The geometry of the pulse varies from a spherical profile to a slab profile, details will be given in ref.25.}
\end{figure}
\newpage
\begin{figure}[t]
\centerline{
\psfig{figure=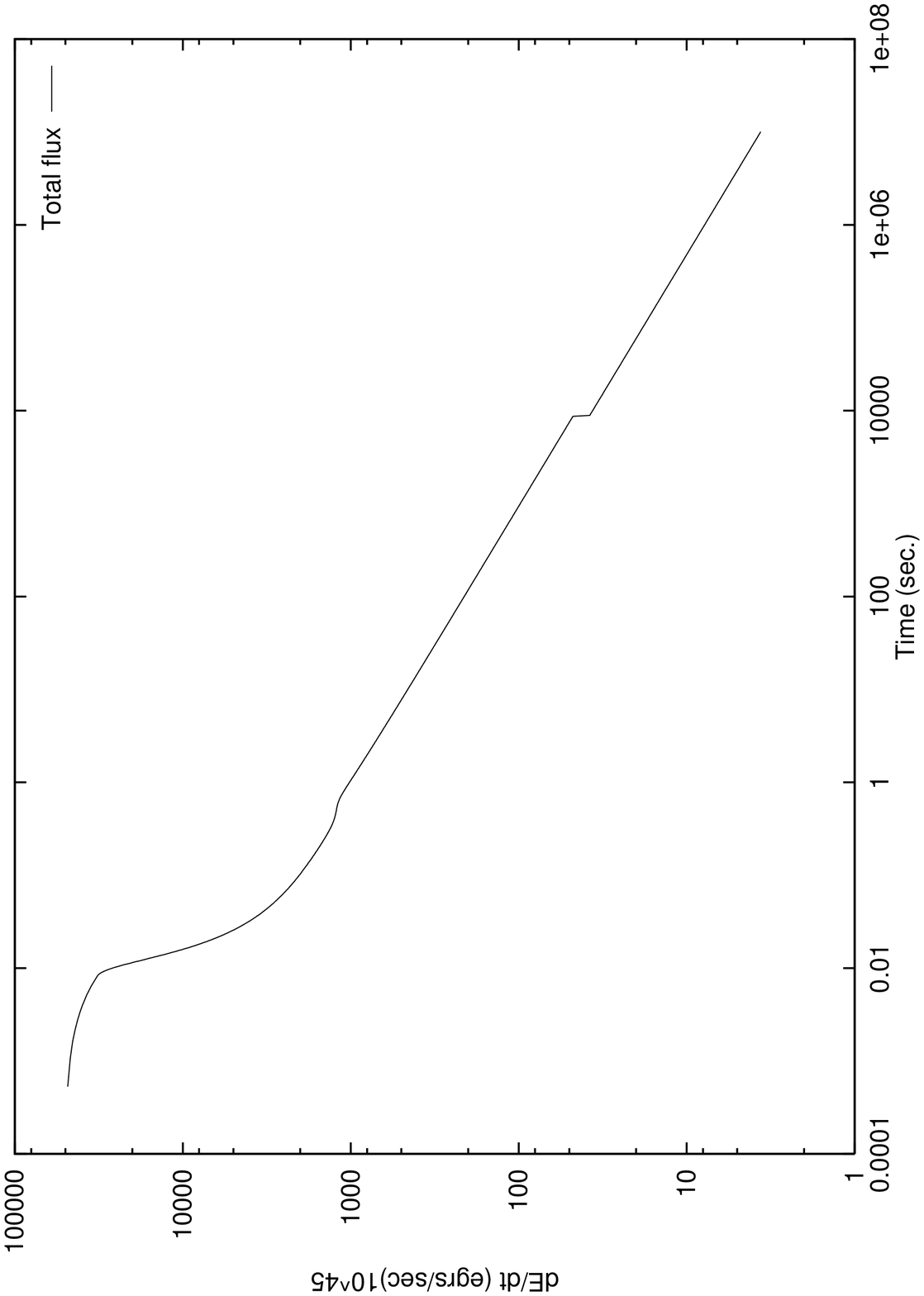}
}
\caption[]{ The energy flux of a PEM-pules from a $10M_\odot$ EMBH with $\xi=0.1$ and with a baryonic matter content of $\rho_b=b\rho_{e^+e^-},b=10^{-4}$. The power law behaviour characterizing the afterglow of Gamma-ray burst up to $10^7$ sec. are recovered\cite{[25]}.}
\end{figure}


\newpage

\begin{references}
\bibitem{[1]} 
Heisenberg W.~and Euler H., {\it Zeits.~Phys.} 
{\bf 69}(1931) 742.

\bibitem{[2]} 
R. Ruffini,"on the Energetics of Black Holes"in "Black Holes" Ed. C. and B.S. de Witt Gordon and Breach N.Y. Paris London 1973;
see also {\it Proceedings of the sixteenth Solvay Conference on Physics at the University of Bruxelles, September, 1973}, Editions de l'Universite' de Bruxelles, 1974.

\bibitem{[3]} 
Presented at the 1972 Texas meeting in N.Y. 1972;
see also R.Leach and R. Ruffini, "On the masses of X-ray sources," {\it Astrophys. J. Lett.} {\bf 180}, L15 (1973).

\bibitem{[4]} 
C. Rhoades, Jr. and R. Ruffini, "Maximum mass of a neutron star." {\it Phys.
Rev. Lett.} 32, 3 (1974).

\bibitem{[5]}
E. Costa {\it et al.}, {\it Nature} {\bf 387}(1997) 783.

\bibitem{[6]}
J. Van Paradijs, {\it et al.} {\it Nature} {\bf 386}(1997) 686.

\bibitem{[7]}
S. R. Kulkarni, {\it et al.} {\it Nature} {\bf 393} (1998) 35

\bibitem{[8]}
J.P. Halpern {\it et al.} {\it Nature} {\bf 393} (1998) 41.


\bibitem{[9]}
A.N. Ramaprakash {\it et al.~}{\it Nature} {\bf
393} (1998) 43.


\bibitem{[10]}
See e.g. M. Rees, R. Ruffini and J.A.Wheeler, {\it Black Holes, Gravitational Waves and Cosmology,} Gordon and Breach, New York, 1974;\\
{\it Neutron Star, Black Holes and Binary X-Ray Sources,} H. Gursky and R. Ruffini, eds. and coauthors, H. Reidel, Amsterdam, 1975;\\
"{\it Physics an Astrophysics of Neutron Star and Black Holes: Proceedings of the LXV International "Enrico Fermi" Varenna Summer School of 1975,} R. Giacconi and R. Ruffini, eds. and coauthors, North Holland, Amsterdam, 1978.

\bibitem{[11]}
See e.g. H. Sato and R. Ruffini, {\it Black Holes. Ultimate State of Stars and General Relativity, Chuo Koron Sha-Tokyo, 1976 (in Japanese} and L.Z. Fang
and R. Ruffini , {\it Basic Concepts in Relativistic Astrophysics}, Shanghai
Science and technology Press, Shanghai, 1981 (in Chinese); English translation by World Scientific, 1983.

\bibitem{[12]}
R. Ruffini in the "Festschrift in honour of the 65th birthday of Riccardo Giacconi" Ed Fang Li Zhi and R. Ruffini . Pub. World Scientific Singapore 1998 -in the press

\bibitem{[13]}
R. Ruffini and J.A. Wheeler, {\it Physics Today} {\bf
January} (1971) 178.


\bibitem{[14]}
B. Carter, in Proceedings of the Eighth Marcel Grossmann Meeting, ed.~T.~Piran, World Scientific, Singapore 1988.


\bibitem{[15]}
D. Christodoulou and R. Ruffini, {\it Phys.~Rev.} {\bf D4}(1971) 3552.

\bibitem{[16]}
R.~Penrose, {it Nuovo Cimento, Rivista} {\bf Vol.~1} {\it special issue}, 1969; 
The first specific example of energy extraction see R.~Ruffini and J.A.~Wheeler, as quoted in D.~Christodoulou {\it Phys.~Rev.~Lett.} {\bf 25}
(1970) 1596; 
See also  R.~Penrose and R.~Floyd, {\it Nature} {\bf 229} (1971) 193.

\bibitem{[17]}
J. Schwinger, {\it Phys.~Rev.} {\bf 82} (1951) 664.

\bibitem{[18]}
T. Damour and R. Ruffini, {\it Phys.~Rev.~Lett.} {\bf
35} (1975) 463.

\bibitem{[19]}
T. Damour and R. Ruffini, in preparation.

\bibitem{[20]}
R. Ruffini, in ``{\it Physics and Astrophysics of Neutron
Stars and Black Holes}", ed.~R.~Giacconi and R.~Ruffini,
North-Holland, Amsterdam, 1978.


\bibitem{[21]}
R. Jantzen and R. Ruffini, submitted for publication.

\bibitem{[22]}
G. Preparata, R. Ruffini and S.-S. Xue, submitted to 
{\it Phys.~Rev.~Lett.}, 1998.

\bibitem{[23]}
G. Preparata, R. Ruffini and S.-S. Xue, submitted to  
{\it Astronomy and Astrophysics Letter}, 1998.

\bibitem{[24]}
R. Ruffini, J. Salmonson, J. Wilson and S.-S. Xue, submitted for publication to AP.J. (1998).

\bibitem{[25]}
R. Ruffini and S.-S. Xue, in preparation, 1998.

\bibitem{[26]}
J. Wilson, {\it Annals of the New York Academy of Sciences} {\bf Vol.~262} (1975) 123.

\bibitem{[27]}
J. Wilson,	{\it Proc.~of the First Marcel Grossmann Meeting on General Relativity}, ed.~R.~Ruffini, North-Holland Pub.~Amsterdam, 1977, p.\ 393.

\bibitem{[28]}
R. Ruffini and T. Treves, {\it Astrophysics Lett.} {\bf 13} (1973) 109.

\bibitem{[29]}
see e.g., B.~Paczy\'nski, {\it Astrophysics Journal} {\bf 363} (1990) 218;\\
R.~Narayan, B.~Paczy\'nski and T.~Piran, {\it Astrophysics Journal} {\bf 395} (1992) L83;\\
R.~Narayan, T.~Piran and A.~Shemi, {\it Astrophysics Journal} {\bf 379} (1991) L17;\\
P.~M\'esz\'aros and M.J.~Rees, {\it Astrophysics Journal} {\bf 482} (1997) L29 and references therein.

\bibitem{[30]}
M.~Ruffert, H.-Th.~Janka, K.~Takahashi and G.~Sch\"afer, {\it Astronomy and Astrophysics (Letter)} {\bf 319} (1997) 122.

\bibitem{[31]}
see e.g., G.~Cavallo and M.F.~Rees, {\it MNRAS} {\bf 183} (1978) 359;
see however, P.~M\'esz\'aros and M.J.~Rees, {\it Astrophysics Journal} {\bf 397} (1992) 570;\\
B.~Paczy\'nski {\it Astrophysics Journal} {\bf 308} (1986) L51;\\
J.~Goodman {\it Astrophysics Journal} {\bf 308} (1986) L47;\\
P.~M\'esz\'aros, P.~Laguna and M.J.~Rees, {\it Astrophysics Journal} {\bf 415} (1993) 181;\\
P.~M\'esz\'aros, M.J.~Rees and H.~Papathanassiou, {\it Astrophysics Journal} {\bf 432} (1994) 181;\\
R.~Sari and T.~Piran, {\it Astrophysics Journal} {\bf 455} (1995) L143;\\
P.~M\'esz\'aros and M.J.~Rees, {\it Astrophysics Journal} {\bf 430} (1994) L931; {\it ibid} {\bf 476} (1997) 232 and references therein.

\bibitem{[32]}
R.~Ruffini, J.~Salmonson, J.~Wilson and S.-S.~Xue, submitted for publication.

\bibitem{[33]}
M.~Vietri, {\it Astrophysics Journal} {\bf 478} (1997) L9.

\bibitem{[34]}
S.W. Hawking {\it et al.} {\it Nature} {\bf 238} (1974) 30.

\end{references}
\end{document}